\pdfoutput=1

\documentclass[11pt]{article}

\usepackage{acl}

\usepackage{times}
\usepackage{latexsym}
\usepackage{graphicx}
\usepackage{amsmath}

\usepackage[T1]{fontenc}

\usepackage[utf8]{inputenc}

\usepackage{microtype}

\usepackage{inconsolata}
\usepackage{bm}
\usepackage{multirow}
\usepackage{boldline}
\usepackage{booktabs}
\usepackage{colortbl}

\newcommand{\nop}[1]{}

\title{Relational Database Augmented Large Language Model}
\author{Zongyue Qin\footnote{Department of Computer Science, University of California, Los Angeles, USA. Correspondence to: Zongyue Qin \texttt{qinzongyue@cs.ucla.edu}}, Chen Luo\footnote{Amazon, Palo Alto, USA.}, Zhengyang Wang\footnotemark[2], Haoming Jiang\footnotemark[2], Yizhou Sun\footnotemark[1]}

\begin{document}
\maketitle
\begin{abstract}
  Large language models (LLMs) excel in many natural language processing (NLP) tasks. However, since LLMs can only incorporate new knowledge through training or supervised fine-tuning processes, they are unsuitable for applications that demand precise, up-to-date, and private information not available in the training corpora.
  This precise, up-to-date, and private information is typically stored in relational databases. Thus, a promising solution is to augment LLMs with the inclusion of relational databases as external memory. This can ensure the timeliness, correctness, and consistency of data, and assist LLMs in performing complex arithmetic operations beyond their inherent capabilities.
  However, bridging the gap between LLMs and relational databases is challenging. It requires the awareness of databases and data values stored in databases to select correct databases and issue correct SQL queries. Besides, it is necessary for the external memory to be independent of the LLM to meet the needs of real-world applications.
  We introduce a novel LLM-agnostic memory architecture comprising a database selection memory, a data value memory, and relational databases. 
  And we design an elegant pipeline to retrieve information from it. 
  Besides, we carefully design the prompts to instruct the LLM to maximize the framework's potential. 
  To evaluate our method, we compose a new dataset with various types of questions.
  Experimental results show that our framework enables LLMs to effectively answer database-related questions, which is beyond their direct ability. 
\end{abstract}

\section{Introduction}

Large language models (LLMs) have demonstrated impressive ability in various tasks~\cite{brown2020language}.
However, LLMs are susceptible to a phenomenon known as ``hallucination'', wherein they may generate text that is factually inaccurate~\cite{ram2023context,schick2023toolformer,parisi2022talm}.
Furthermore, LLMs are inherently bounded by their training data and must be trained to incorporate new information, rendering them unable to handle tasks that demand up-to-date or private information, which are common in real-world applications. A common solution is to enhance an LLM with a memory module that stores external knowledge~\cite{hu2022empowering,guu2020retrieval,izacard2022few}. It allows the LLM to retrieve relevant information from the memory to generate factual responses. Existing LLM memory modules mainly utilize unstructured or semi-structured knowledge such as raw text and knowledge graphs. However, using relational databases for LLM memory can be a better option because (1) it is easier to maintain the correctness, consistency, and timeliness of the information; (2) it supports complicated arithmetic and logical reasoning over stored data; (3) it is widely used in real-world applications.

Therefore, the goal of this work is to build an external memory based on a collection of relational databases to augment LLMs. 
Moreover, we believe it is necessary to make the external memory LLM-agnostic, which means the overall framework should not modify or fine-tune the LLM. It is because (1) many commercial LLMs can only be accessed via API calls (e.g., ChatGPT, Claude2); (2) most people do not have enough resources to train or fine-tune a LLM, while it is much cheaper to just run a LLM for inference; (3) since newer and more powerful LLMs are constantly emerging, the need to switch the base LLM could be common. So an LLM-agnostic approach can better serve the needs of real-world applications.

To achieve our objective, there are multiple challenges to tackle. First, given a retrieval target described in natural language, how to generate the correct SQL is challenging. In particular, existing text-to-SQL methods~\cite{pourreza2023din,li2023resdsql,kocon2023chatgpt} cannot appropriately handle the discrepancy between how an entity is described in natural language and how it is stored in the database. 
Second, a new retrieval paradigm is needed to enable LLMs to retrieve information from multiple databases, which is fundamentally different from the situation with a single database. Only using SQL query is enough to retrieve information from a specific database. But due to the limitation of SQL syntax, it cannot handle multiple databases, which makes it important to extend the power of SQL.
Third, given candidate databases and a question that might require zero, one, or multiple retrieval steps, it is unclear what is the best way to let the LLM generate a concrete retrieval plan.  

To address the above challenges, we propose a novel LLM-agnostic memory architecture that consists of a database selection memory that helps retrieve relevant databases, a data value memory that helps retrieve relevant database values, and relational databases that stores the information to help generate responses. 
And we design an elegant pipeline to retrieve information from the memory. 
We also propose an overall framework to integrate the LLM and the external memory that lets LLM automatically determine if a retrieval from the memory is necessary and lets LLM utilize the retrieved information to generate responses.
To evaluate our framework, we compose a new dataset using Q\&A pairs from four public datasets. The experiment results show that our method allows the LLMs to answer questions that require accessing database contents, which is beyond the direct ability of LLMs, with reasonable accuracy. And it also slightly improves the accuracy for questions that do not need accessing databases. 
In summary, our work bridges the gap between LLMs and the utilization of relational databases. We believe it would be useful in diverse applications such as virtual assistants and factual Q\&A.

\section{Related Work}

\textbf{Knowledge enhanced Language Model}.
The knowledge enhanced language models equip language models with external knowledge sources as memory. The majority of knowledge enhanced language models focus on utilizing unstructured or semi-structured knowledge such as raw text and knowledge graphs~\cite{guu2020retrieval,izacard2022few,hu2022empowering}. Earlier works change the architecture of LMs or require continual training~\cite{izacard2022few,hu2022empowering}. 
RePlug~\cite{shi2023replug} introduces a framework that treats the language model (LM) as a black box and augments it with a tuneable retrieval model. \cite{ram2023context} introduces the concept of In-Context RALM, which shows considerable potential to increase the prevalence of LM grounding. Existing knowledge enhanced LMs mainly rely on semantic embeddings of documents or entities to retrieve relevant information. Unfortunately, it is difficult to apply semantic retrieval to relational databases since it does not support complicated arithmetic and logical operations that SQL can do.

A concurrent work,  ChatDB~\cite{hu2023chatdb}, shares a similar idea with ours. 
However, it only introduces a general framework that utilizes the LLM to issue SQL queries to interact with a single database. It cannot handle the setting with multiple databases, which is fundamentally more challenging because it requires a new retrieval paradigm that exceeds the capability of SQL. In addition, it does not consider any challenges discussed in the introduction.

\textbf{Tool Augmented Language Model}.
Similar to knowledge enhanced LMs, existing works on tool augmented LMs teach the language model to use tools in two different ways: In-context learning and Fine-tuning~\cite{li2023api}. The former includes the instructions and the examples of all the candidate tools in the prompt, which do not train the model at all but is limited by the context length. In comparison, the latter~\cite{schick2023toolformer,parisi2022talm} is fine-tuning the language model in supervised or self-supervised way, which has no length problem but might damage the robustness of the model. However, the tools considered in these studies are simple (e.g., search engines, calculators, calendars. In comparison, relational databases are much more complicated. Any people can easily use search engines and calculators, but only trained SQL experts can retrieve information from relational databases correctly. 

\textbf{Text-to-SQL Model}.
The earlier works in text-to-SQL task learn a sequence-to-sequence model to encode a given natural language question with the database schema and leverage a decoder to predict the target SQL~\cite{li2023resdsql,zhao2022importance,scholak2021picard}. Recent studies~\cite{pourreza2023din,dong2023c3} show that LLMs achieve state-of-the-art accuracy in text-to-SQL tasks by utilizing in-context learning and attain leading positions in text-to-SQL benchmarks. 
However, existing models lack a viable solution to handle the database value unawareness problem. That is, the values in the natural language question might differ from the values stored in the database. Not knowing what values are stored, the model would generate reasonable but false SQL queries. 
Existing solutions of this problem include fine-tuning the LM on the specific database or inputting all the values in the selected columns to the LM~\cite{li2023resdsql}. However, neither solution is ideal because (1) fine-tuning cannot handle new databases or databases that are constantly updating; 
and (2) input all database values is only feasible for tiny databases with hundreds of cells and poses a great risk of leaking private information. 

\section{Framework Overview}

Here we first formulate our problem. Then, we give a high-level description to our framework. 

\subsection{Problem Definition}

Our method aims to retrieve necessary information from a collection of relational databases to help LLMs in producing accurate and database-specific outputs. In the conventional LM paradigm, both input and output consist of text. However, a common use case of relational databases involves users inputting natural language questions to retrieve specific data from the database systems. So in this work, given the user input, which \textbf{may or may not} require information from relational databases, the expected output could be \textbf{either a textual string or a result in the form of SQL query result}.

\subsection{Framework Modulization}

\begin{figure}[t]
    \centering
    \includegraphics[width=0.92\linewidth]{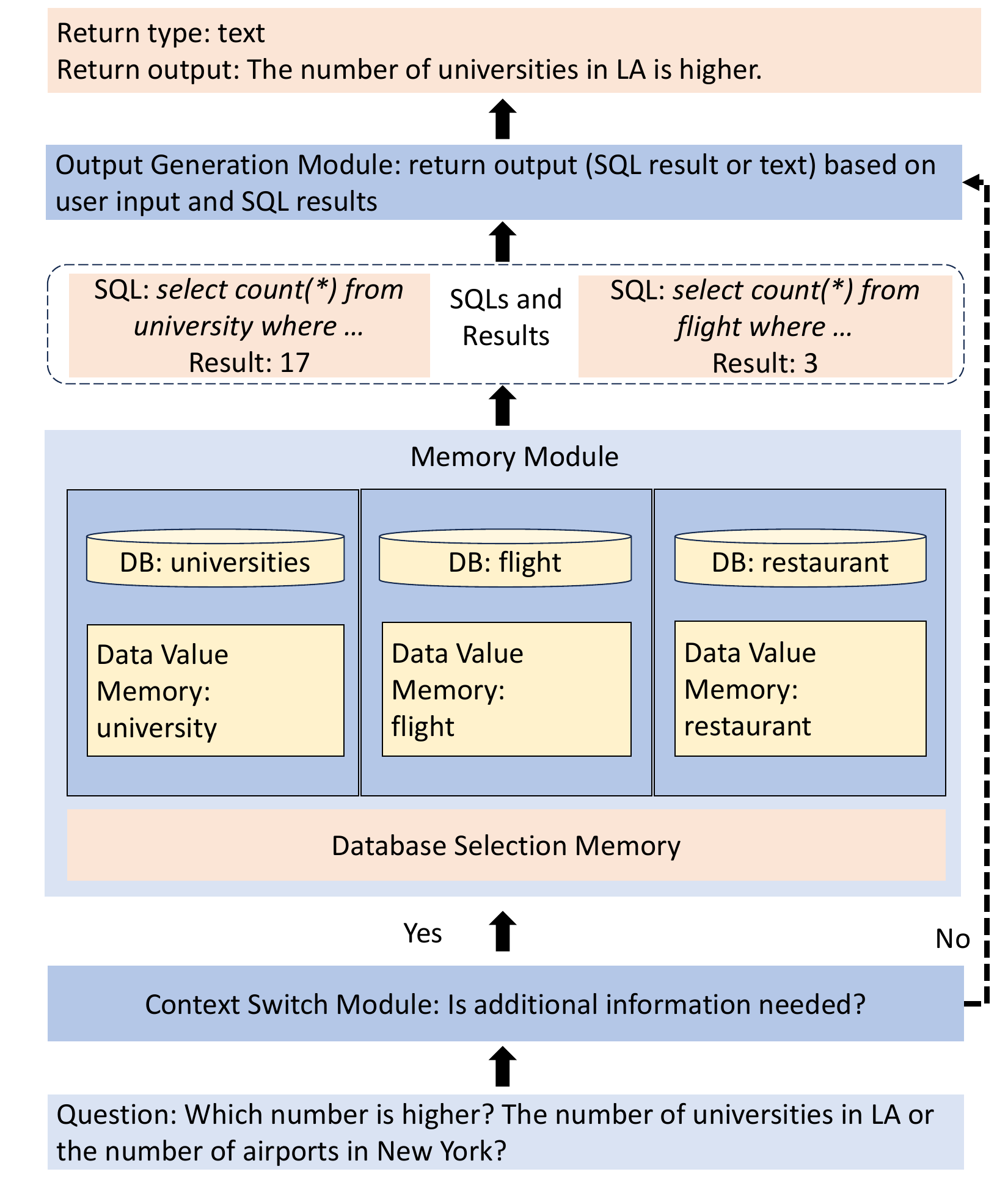}
    \caption{Overview of our framework. The context switch module first determines if additional information is needed. The memory module retrieves the information from a collection of databases. The output generation module returns the final response based on the user input and the retrieved information (if available). 
    }
    \label{fig:overview}
    \vspace{-3mm}
\end{figure}

Figure \ref{fig:overview} illustrates the overall framework. Our framework uses prompts to instruct the LLM to complete some steps (e.g., context switch module and output generation module). All the prompts of our method are included in the appendix in the supplementary materials.

\begin{figure}
    \centering
    \includegraphics[width=0.92\linewidth]{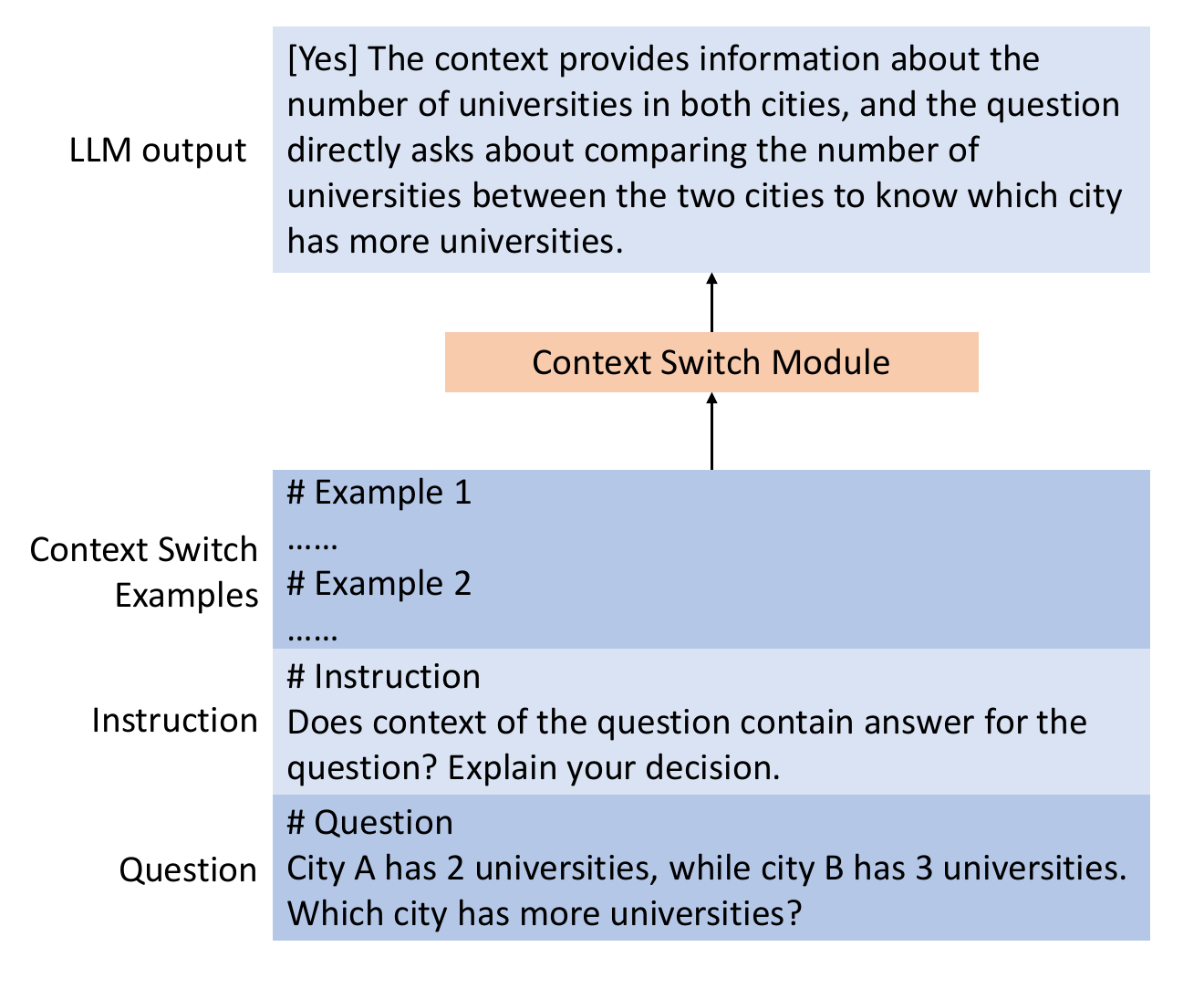}
    \caption{An example showing the input and output of the context switch module 
    \vspace{-3mm}
    }
    \label{fig:context_switch}
\end{figure}

\textbf{Context Switch Module}.
The goal of this module is to reduce unnecessary information retrieval.
Given an input, retrieval should be avoided if the input already provides enough information in the context. 
So the context switch module determines if a question is answerable based on the input context, which is a classic task for LM~\cite{rajpurkar2018know}. 
Previous studies have shown that LLMs could address the task with reasonable accuracy~\cite{kocon2023chatgpt}. So we implement this module by calling the LLM with our designed prompts. An example of the prompt and how the context switch works is shown in Figure \ref{fig:context_switch}.

\begin{figure}
    \centering
    \includegraphics[width=0.92\linewidth]{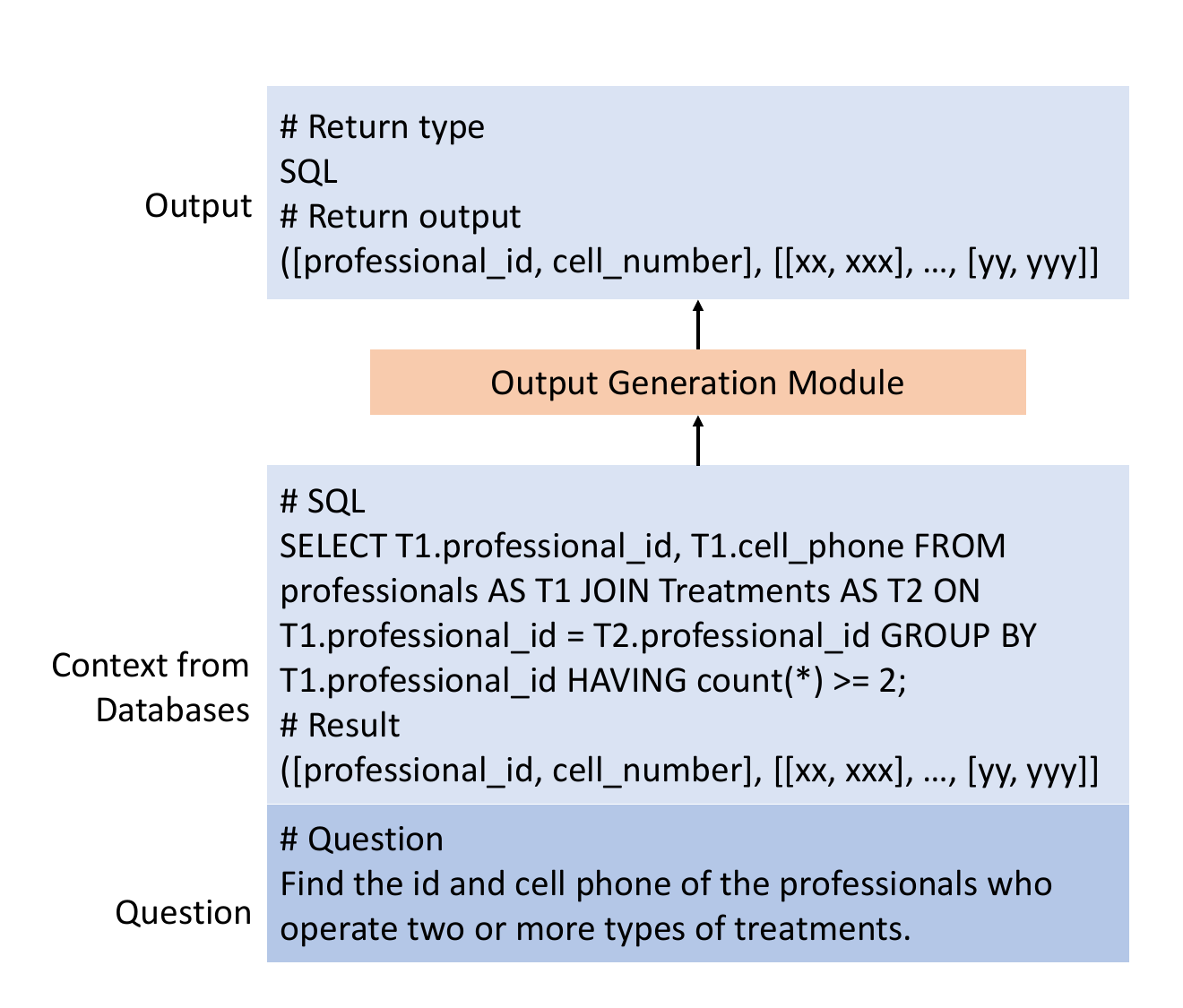}
    \caption{An example showing the input and output of the output generation module.}
    \label{fig:output}
    \vspace{-6mm}
\end{figure}

\textbf{Memory Module}.
Here we only describe the functionality of the memory module. The details are introduced in Section \ref{sec:memory}.
When the input context does not contain sufficient information, the memory module is activated to retrieve relevant information from relational databases. The memory module returns a list of SQL results. It is possible that one of them could be directly returned as the final response. For example, if the input question is "\textit{Show me all the Thai restaurants in New York}", the memory module might return a list of restaurant names as one of the SQL results.  
In addition, the retrieval might fail and return an empty list. 
It can happen if the databases do not contain any relevant information or there is an error during the retrieval process.
In this case, the output generation module will let the LLM directly answer the question based on its own knowledge. 

\textbf{Output Generation Module}.
After retrieving information from the databases, the output generation module generates outputs based on the user inputs and the retrieved SQL results. 
Conventionally, the retrieved information is directly concatenated with the user input and fed to the LLM to generate the response~\cite{hu2022empowering,izacard2022few,guu2020retrieval}. 
However, since the output of our framework could be either SQL results or text strings, an extra step is needed to determine the response type. 
In particular, the LLM should return a SQL result if its corresponding SQL query is semantically equivalent to the user input. 
Therefore, the output generation module calls the LLM to check if there is any SQL query semantically equivalent to the user input. If there is, then the corresponding SQL result is directly returned by the module. Otherwise, the module concatenates the SQL queries, the SQL results, and the user input together, and lets the LLM generate the response. 
An example of this module is shown in Figure \ref{fig:output}. 

\section{LLM-agnostic Memory\label{sec:memory}}

\begin{figure}
    \centering
    \includegraphics[width=0.95\linewidth]{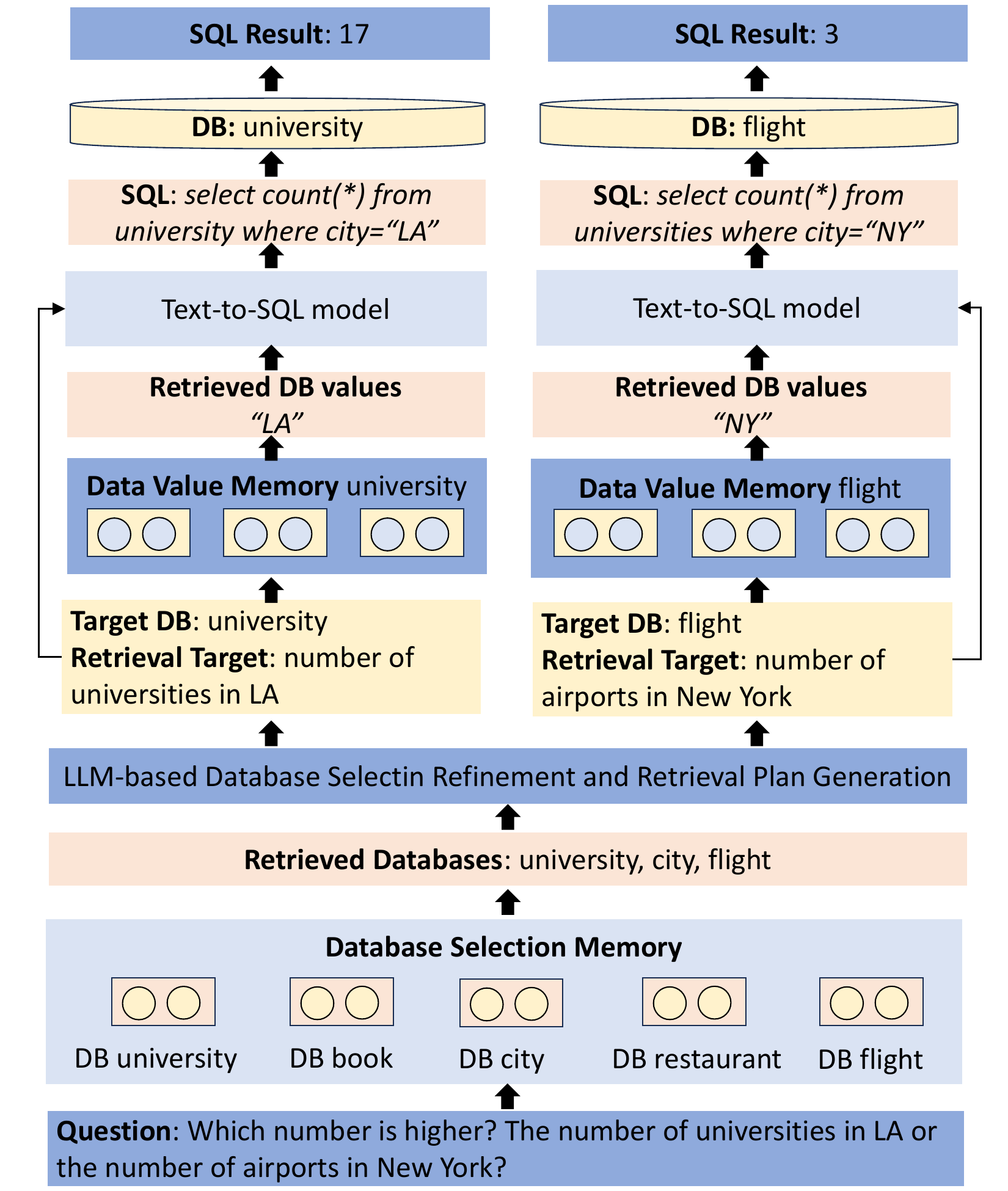}
    \caption{Illustration of the proposed memory module. The database selection memory returns top K relevant databases. Then the LLM refines the database selection and generates a retrieval plan following our proposed strategy. For each retrieval target, the data value memory returns relevant data values from the database to help generate correct SQL queries. Then information is retrieved using SQL. 
    }
    \label{fig:memory}
\end{figure}

As shown in Figure \ref{fig:overview}, our proposed memory module mainly consists of a database selection memory, a data value memory, and the databases. Figure \ref{fig:memory} illustrates how our proposed memory works in detail. To retrieve information, the database selection memory first retrieves relevant databases. Next, the LLM refines the database selection and makes a concrete retrieval plan for each candidate database. Then, the data value memory retrieves database values relevant to the user input, which are then used to help generate accurate SQL queries. Finally, the SQL queries are issued to the corresponding databases to retrieve information. All the prompts used in the memory module are shown in the appendix in the supplementary material.


\subsection{Database Selection Memory}

The first step for the memory module is to select which databases might contain information relevant to the inputs. The selection should (1) return databases with contents that are semantically relevant to the input questions, (2) be efficient enough to handle a large number of databases, and (3) easily handle newly added databases.

So we propose an embedding-based database selection memory to satisfy the above requirements. It employs an independent embedding function to convert databases and input questions into embeddings such that their semantic relevance is reflected by their proximity in the embedding space. The retrieval can be done efficiently with nearest neighbor search algorithms~\cite{malkov2018efficient,johnson2019billion}. Besides, since the embedding is obtained via an independent function, the database selection memory can handle newly added databases and satisfies the LLM-agnostic principle.

\nop{
\begin{figure}
    \centering
    \includegraphics[width=0.9\linewidth]{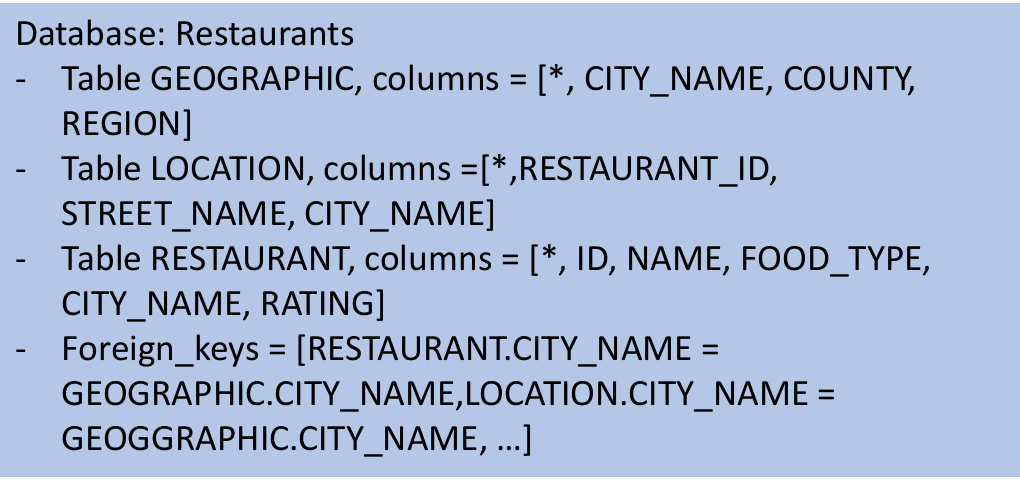}
    \caption{An example of database schema}
    \label{fig:schema}
    \vspace{-3mm}
\end{figure}
}

To obtain such an embedding function, we choose to fine-tune an independent language model (LM) due to its strength in capturing semantic similarity and good generalization ability. Since LMs take text strings as inputs, we convert database schemas into text strings as inputs to the embedding function, following the conventions in previous text-to-SQL studies~\cite{pourreza2023din}. 
Given a user input $\bm{x}$ and a database $D$, predicting if a database is relevant to an input question is a binary classification task. The output is a probability score $p_{\bm{x},D}$ between 0 and 1. So the model can be trained with binary cross-entropy loss.
\begin{equation}
    -y_{\bm{x},D}\log p_{\bm{x},D} - (1-y_{\bm{x},D})\log(1-p_{\bm{x},D})
\end{equation}

However, there is no public dataset with such labels. So we compose a new training dataset by utilizing two public text-to-SQL datasets, Spider~\cite{yu2018spider} and Dr Spider \cite{chang2023dr}. The original datasets consist of a list of (question, SQL, database) triplets, which can be used to obtain the positive labels. We generate negative labels by replacing the ground-truth database with a random one. In addition, to mimic real-world questions requiring accessing multiple databases, we concatenate $K$ random questions together as a composite question. 
With the new dataset, we can fine-tune the LM in a supervised way.


\subsection{LLM-based Database Selection Refinement and Retrieval Plan Generation\label{sec:plan}}

A straightforward way to generate a retrieval plan is to let the LLM generate a retrieval target for each database returned by the database selection memory. 
However, it does not consider the problem caused by the false positive databases returned by the database selection memory. Database selection memory is prone to return false positive databases because some databases contain keywords relevant to the input questions, while other databases might contain more accurate and comprehensive information. 
For example, assume there are two databases ``city'' and ``hospital''. The city database contains some general information for each city, while the hospital city records which city each hospital is located in. Given a question ``\textit{How many hospitals are located in city Seattle}'', database selection memory might determine both databases are relevant to the question since they contain keywords "hospital" and "city", respectively. But we know the hospital database can provide accurate information to the question, which makes the city database redundant.
The redundant retrievals will harm the overall response efficiency. More importantly, the information retrieved from the false positive databases might be inaccurate and cause the LLM to generate wrong responses.

To handle this problem, we propose an elegant strategy that utilizes the dynamic planning ability of the LLM to refine database selection and generate concrete retrieval plans. Specifically, the LLM takes the question and all the candidate database schemas as input and is instructed to generate the retrieval targets from the most relevant databases to the least relevant ones until it gets all the information needed. This approach avoids the redundant retrieval in the previous example because the LLM first selects the hospital database and generates a retrieval target. Then it realizes it already has all the information, so it will ignore the city database.

Notice that it is infeasible to select databases using our database selection refinement strategy without the database selection memory. This is due to the length constraint of the input prompts. Since the refinement strategy requires the input prompt to include all database schemas, and each database schema usually contains hundreds of tokens, the number of databases our refinement strategy can handle is limited. But the database selection memory addresses this issue by only returning the top-K most relevant databases.

\subsection{Data Value Memory for SQL Generation}

Given a database and a retrieval target, the next step is to parse the target into a SQL query. 
To write a correct SQL, a human expert has to know (1) the database schemas (e.g., table and column names), (2) how to express the logic of the retrieval target into SQL, and (3) the data values stored in the database. Previous studies on text-to-SQL~\cite{pourreza2023din,dong2023c3,li2023resdsql} mainly focus on the first two parts but lack practical solutions for the discrepancy between how an entity is described in natural language and how it is stored in the database.
For example, assume there is a table of restaurants where the city Los Angeles is stored as ``LA''. Given a question ``\textit{show me the restaurant located in Los Angeles}'', the correct SQL should be ``
\textit{SELECT name FROM restaurant WHERE location=‘LA’}
''. But if a text-to-SQL model is unaware of the values stored in the table, it is likely to generate ``
\textit{SELECT name FROM restaurant WHERE location=‘Los Angeles’}
''. 
Common solutions to this problem include fine-tuning the model on the database or inputting the entire database into the model. The former solution cannot handle newly added databases and the latter one is only feasible for tiny databases.

So we propose a data value memory to address the problem. It returns the database values that are relevant to the input question, which are then fed to the text-to-SQL model to help generate correct SQL queries.
As shown in Figure \ref{fig:memory}, each database has its own data value memory. Each data value memory consists of multiple column-wise memories, where each column-wise memory stores the semantic embeddings of data values in a column. 

To retrieve database values relevant to the input question, the data value memory first calls an existing text-to-SQL method to generate a candidate SQL query in a database-value-unaware way. Then, a SQL parser identifies the columns and values that appeared in the conditions of the candidate SQL query (e.g., 
[restaurant.location, ‘Los Angeles’] in the previous example). 
Next, the module searches the corresponding column-wise memory to find possible value synonyms of the identified value based on their semantic embeddings. 

To make the data value memory LLM-agnostic and capable of handling new databases, we use an independent LM to generate the semantic embeddings for database values and the identified values. Notice that it is unnecessary to fine-tune the LM since the goal is to find synonyms and the sentence embeddings of a well-trained language model should already capture the semantic similarity~\cite{reimers2019sentence}.   

Given the returned database values and the candidate SQL, the memory module instructs the LLM to identify if there is any error in the candidate SQL and correct it. For example, given the candidate SQL, ``
\textit{SELECT name FROM restaurant WHERE location=‘Los Angeles’}
'' and the database value ''LA'', the LLM can identify the error in the where clause and correct it. 
After the SQL queries are generated, they are sent to the corresponding databases for execution. The returned SQL results are sent to the output generation module.

\section{Experiments}


To evaluate our framework, we compose a new dataset by collecting questions from four public datasets: SQUAD-v2~\cite{rajpurkar2018know}, Commonsense QA~\cite{talmor2018commonsenseqa}, Spider~\cite{yu2018spider}, and Dr-Spider~\cite{chang2023dr}. There are three types of questions in our dataset: 
(1) questions do not require accessing databases, (2) questions require accessing one and only one database, and (3) questions require accessing more than one database. 
The details of how the dataset is collected are described in the appendix. Figure \ref{fig:dataset} shows the detailed composition of the dataset.


\begin{figure}
    \centering
    \includegraphics[width=0.99\linewidth,height=0.1\textheight]{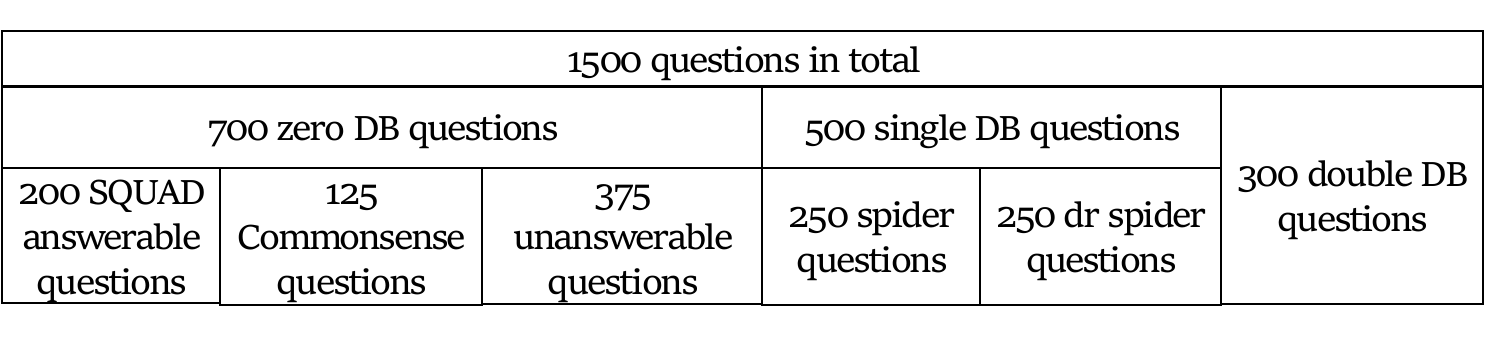}
    \vspace{-3mm}
    \caption{Dataset Composition}
    \label{fig:dataset}
\end{figure}


We implement our framework with \textit{GPT-3.5-turbo-16k-0613}~\cite{brown2020language} and \textit{claude-2}~\cite{Anthropic_2023} as the base LLM, respectively. Both models are commercial language models and are accessed via APIs. 
We use DIN-SQL~\cite{pourreza2023din}, which is a state-of-the-art LLM-based text-to-SQL model, to generate candidate SQL queries.
The base LLMs are used as the basis of our framework and the text-to-SQL model. In addition, we use bert-large-uncase LM as the independent LM for database selection memory and datab value memory. The database selection memory returns the top 5 candidate databases and the datab value memory returns the top 10 database values for each column.

\subsection{SQL and Answer Accuracy}
First, we conduct experiments to answer the following research questions. \textbf{RQ1}: If our method can retrieve correct information; \textbf{RQ2}: If our method can let LLMs generate correct responses; \textbf{RQ3}: Does our method affect LLMs' performance on questions not require accessing databases.   

\begin{table*}[]
    \small
    \caption{SQL and answering accuracy for the LLM-only method and our proposed method. Bold numbers mark the highest performance for the same foundation model. Shaded columns (Single DB, Double DM) mark the questions requiring database access. 
    }
    \label{tab:acc}
    \centering
    \begin{tabular}{c|c|c|>{\columncolor[gray]{0.9}}c>{\columncolor[gray]{0.9}}c|>{\columncolor[gray]{0.9}}c>{\columncolor[gray]{0.9}}c}\hlineB{3}
       \multirow{2}{*}{Setting}  & \multirow{2}{*}{Method}& Zero DB&
       \multicolumn{2}{>{\columncolor[gray]{0.9}}c|}{Single DB} & \multicolumn{2}{>{\columncolor[gray]{0.9}}c}{Double DB} \\ \cline{3-7}
        &  & Answer Acc& SQL Acc & Answer Acc & SQL Acc & Answer Acc \\\hlineB{3}
    GPT3.5, & LLM only  & 0.78 & 0 & 0 & 0 & 0\\
     $|\mathcal{D}|=5$& ours  &  \textbf{0.80} & \textbf{0.64} & \textbf{0.63} & \textbf{0.60} & \textbf{0.37}\\
    \hlineB{3}
    GPT3.5, & LLM only  & 0.78  & 0 & 0 & 0 & 0\\
     $|\mathcal{D}|=20$& ours  & \textbf{0.79} & \textbf{0.62} & \textbf{0.60} & \textbf{0.55} & \textbf{0.32}\\
    \hlineB{3}
     Claude-2 & LLM only &\textbf{0.87} & 0 & 0& 0& 0 \\
     $|D|=5$& ours& 0.86 & \textbf{0.62} & \textbf{0.61}& \textbf{0.48}& \textbf{0.43}\\
     \hlineB{3}
     Claude-2 & LLM only &0.87& 0 & 0& 0& 0 \\
     $|D|=20$& ours& \textbf{0.91} & \textbf{0.66} & \textbf{0.54}& \textbf{0.54}& \textbf{0.44}\\
     \hlineB{3}
    \end{tabular}
\end{table*}

\begin{table*}[]
    \small
    \caption{Effect of datab value memory (DVM) on SQL and answering accuracy. 
    }
    \label{tab:acc_dvm}
    \centering
    \begin{tabular}{c|c|c|>{\columncolor[gray]{0.9}}c>{\columncolor[gray]{0.9}}c|>{\columncolor[gray]{0.9}}c>{\columncolor[gray]{0.9}}c}\hlineB{3}
       \multirow{2}{*}{}  & \multirow{2}{*}{}& Zero DB&
       \multicolumn{2}{>{\columncolor[gray]{0.9}}c|}{Single DB} & \multicolumn{2}{>{\columncolor[gray]{0.9}}c}{Double DB} \\ \cline{3-7}
        & & Answer Acc & SQL Acc & Answer Acc & SQL Acc & Answer Acc \\\hlineB{3}
    \multirow{2}{*}{GPT-3.5} & no DVM & 0.77 & 0.61 & 0.61 & 0.50 & \textbf{0.37} \\
    & with DVM  & \textbf{0.80}  & \textbf{0.64} & \textbf{0.63} & \textbf{0.60} & \textbf{0.37}\\
    \hlineB{3}
     \multirow{2}{*}{Claude-2} & no DVM  & 0.79 & 0.51 & 0.50& 0.47& \textbf{0.43}\\
     & with DVM & \textbf{0.86}  & \textbf{0.62} & \textbf{0.61}& \textbf{0.48}& \textbf{0.43}\\
     \hlineB{3}
    \end{tabular}
\end{table*}

\textbf{Metrics}. We define two metrics, SQL accuracy and answer accuracy, for our experiments. The SQL accuracy evaluates if a model can retrieve the correct information from databases. The metric is only defined for questions that require accessing databases. The answer accuracy measures if the returned output is correct. It is defined for all types of questions. The specific definitions of SQL accuracy and answer accuracy for each question type are given below. Notice that unanswerable questions are not included in this evaluation because they do not have ground-truth answers.

\begin{itemize}
    \item \textbf{\textit{Zero DB Questions}}. These questions do not require accessing any database. Given the ground-truth answer and the answer returned, we use the GPT-4 model~\cite{gpt4} to determine if the two answers are equivalent. The answer accuracy is 1 if the GPT-4 model returns true, otherwise, it is 0.
    \item \textbf{\textit{Single DB Questions}}. They are sampled from Spider and Dr. Spider datasets. Each question can be parsed into a single SQL query, so the ground-truth answer is always a SQL result. For a single DB question, the SQL accuracy is 1 only if at least one of the SQL results returned by the text-to-SQL and retrieval module exactly matches the ground-truth SQL result. Otherwise, the SQL accuracy is 0. The answer accuracy is 1 only if the output generation module returns a SQL query result and it exactly matches the ground-truth SQL result. Otherwise, the answer accuracy is 0.
    \item \textbf{\textit{Double DB Questions}}. They require the LLM to issue two SQL queries to two different databases and to process the SQL results to generate the final response in text format. For a double DB question, the SQL accuracy is 1 only if for each of the two ground-truth SQL results, there is an exact match in the SQL results returned by the text-to-SQL and retrieval module. Otherwise, the SQL accuracy is 0. The answer accuracy is 1 only if the SQL accuracy is 1 and the answer is equivalent to the ground-truth answer, which is determined by the GPT-4 model.
\end{itemize}

\textbf{Results}.
We compare our method with calling LLM to answer the question in an end-to-end way, which is denoted as "LLM only". 
For each base LLM, we conduct two sets of experiments with the number of total databases ($|\mathcal{D}|$) being 5 and 20, respectively. Table \ref{tab:acc} shows the results. First, we can see that the LLM-only method always has a SQL and answer accuracy of 0 for single and double DB questions. This is because the LLM-only method cannot issue any SQL queries to retrieve the information needed to answer the question. So it either returns a sentence indicating it does not know the answer or returns responses inconsistent with the information stored in the databases. Meanwhile, our proposed method has a reasonable SQL and answer accuracy on these questions. The SQL accuracy of our method for single-database and double-database questions indicates our method can retrieve information from relational databases effectively. The fact that the answer accuracy and SQL accuracy are close in the single database questions suggests the output generation module effectively predicts what type of responses should be returned. However, for double database questions, there is a gap between the SQL accuracy and answer accuracy. It is because when provided with correct numbers, LLMs cannot correctly compute arithmetic results due to their inherent limitation. 

In addition, our method answers zero DB questions with higher accuracy in three out of four cases. It shows that even though our method is not specifically designed for zero DB questions, the CoT prompts used in the context switch module and the output generation module help improve LLM's effectiveness on conventional Q\&A questions.

Interestingly, Claude-2 has a higher answer accuracy when the total number of databases increases from 5 to 20. We think it might be caused by the composite interaction between Claude-2 and the database selection memory. For example, originally there is a database A that is falsely selected by the database selection memory and Claude-2 also fails to prune it during database selection refinement. But when new databases are added, the database selection memory returns database B instead of A because B has a higher relevance to the input question. And Claude-2 might successfully prune B the database selection memory and the base LLM are independent. As the result, the database selection accuracy might increase when new databases are added, which in turn improves the SQL and answer accuracy.

\subsection{Effect of Database Selection Refinement}

Table \ref{tab:db_pruner} shows the effect of the database selection refinement strategy (Sec \ref{sec:plan}) with a total of 20 candidate databases. 
For each candidate database, whether to make a retrieval from it is a binary classification problem. So we report the precision, recall, and f1-score of the database selection. We can see that the precision without refinement is low. It empirically proves our claim that false positive databases are common. 
Meanwhile, the refinement strategy helps the LLM to achieve a significantly higher precision and f1-score. It proves our strategy can improve the effectiveness of database selection. 

\begin{table}[]
    \footnotesize
    \caption{Effect of Database Selection Refinement.}
    \label{tab:db_pruner}
    \centering
    \begin{tabular}{c|c|c|c}\hlineB{3}
         & precision & recall & f1-score \\\hlineB{3}
    no refinement& 0.13 & \textbf{0.89} & 0.23\\
    \rowcolor[gray]{0.9}with refinement & \textbf{0.64} & 0.81 & \textbf{0.71}\\\hlineB{3}
    \end{tabular}
\end{table}

\nop{
\subsubsection{Context Switch}

Next, we show the effect of the context switch module on the overall agent design. Since the goal of context switch control is to avoid unnecessary retrieval, we evaluate its effect on the database selection to see if it could prevent false positive database retrieval.
Table \ref{tab:context_switch} shows the precision, recall, and f1-score of database selection where each question has 5 candidate databases. We can see that adding the context switch improves the precision by 0.05 without decreasing the recall. It suggests that the context switch successfully prevents redundant database retrieval and has no negative effect on the database selection recall. 

\begin{table}[]
\footnotesize
    \caption{Effect of context switch to the effectiveness of database selection}
    \label{tab:context_switch}
    \centering
    \begin{tabular}{c|c|c|c}\hlineB{3}
         & precision & recall & f1-score \\\hlineB{3}
    without Context Switch & 0.61 & \textbf{0.90} & 0.73\\
    with Context Switch & \textbf{0.66} & \textbf{0.90} & \textbf{0.76}\\\hlineB{3}
    \end{tabular}
\end{table}
}

\subsection{Effect of Data Value Memory}

\begin{table}[]
    \caption{Effect of datab value memory (DVM) to text-to-SQL accuracy}
    \small
    \label{tab:value_retriever}
    \centering
    \begin{tabular}{c|c|c}\hlineB{3}
     perturbation type & no DVM & with DVM \\\hlineB{3}
      \rowcolor[gray]{0.9} value synonym & 0.49 & \textbf{0.62}\\
      column value & 0.67 & \textbf{0.71}\\
      column synonym & \textbf{0.56} & \textbf{0.56}\\
      keyword carrier & 0.83 & \textbf{0.85}\\
      column attribute & 0.55 & \textbf{0.58}\\
      keyword synonym & \textbf{0.61} & \textbf{0.61}\\
      column carrier & \textbf{0.63} & 0.57\\
      multi-type & 0.55 & \textbf{0.58}\\
      others & \textbf{0.73} & \textbf{0.73}\\\hlineB{3}
    \end{tabular}
\end{table}

Next, we evaluate the effect of datab value memory. Table \ref{tab:acc_dvm} shows the SQL and answer accuracy of our method with and without datab value memory. We can see that adding datab value memory increases the SQL accuracy by an average of 0.063 and increases the answer accuracy by an average of 0.038. 

In addition, we use a subset of questions in the Dr-spider dataset to evaluate the effect of the datab value memory on the robustness of text-to-SQL parsing. We choose the natural language question perturbation subset of the Dr-spider dataset, which modifies the input question with various kinds of paraphrases. Table \ref{tab:value_retriever} shows the exact execution accuracy (i.e., exact match accuracy of SQL result) between generating SQL with and without datab value memory under different kinds of perturbation. The definition of each perturbation type can be found in~\cite{chang2023dr}. First, we can see that under value-synonym perturbation, which is the perturbation datab value memory designed for, it significantly improves the accuracy by 0.13. In addition, we can see that in 8 out of 9 perturbation types, adding the memory results in a higher or equal execution accuracy. Therefore, we conclude that our datab value memory significantly improves the robustness and effectiveness of text-to-SQL models.

\nop{
\subsection{Case Study}

Figure \ref{fig:case_study} shows a sample case of the Claude-2 model on a double database question where 20 candidate databases are provided. The question asks the agent to compare the number of singers and the number of employees. Notably, the agent has to access two databases, "singer" and "employee\_hire\_evaluation" to obtain the actual values of the numbers. From the retrieval plan, we can see that the agent correctly understands the question and generates two retrieval goals to obtain the necessary data. And it correctly selects which databases are relevant and generates correct SQL queries. Furthermore, when provided with the SQL queries and results, the output generation module manages to extract information from them to generate the output. From this example, we can conclude that our agent has the ability to generate retrieval plans based on the provided databases and utilize the SQL results to generate outputs. However, we notice that in other double database questions, the LLM might make wrong answers when comparing the two numbers due to the inherent limitation of LLMs 

\begin{figure}
    \centering
    \includegraphics[width=0.95\linewidth,height=0.6\textheight]{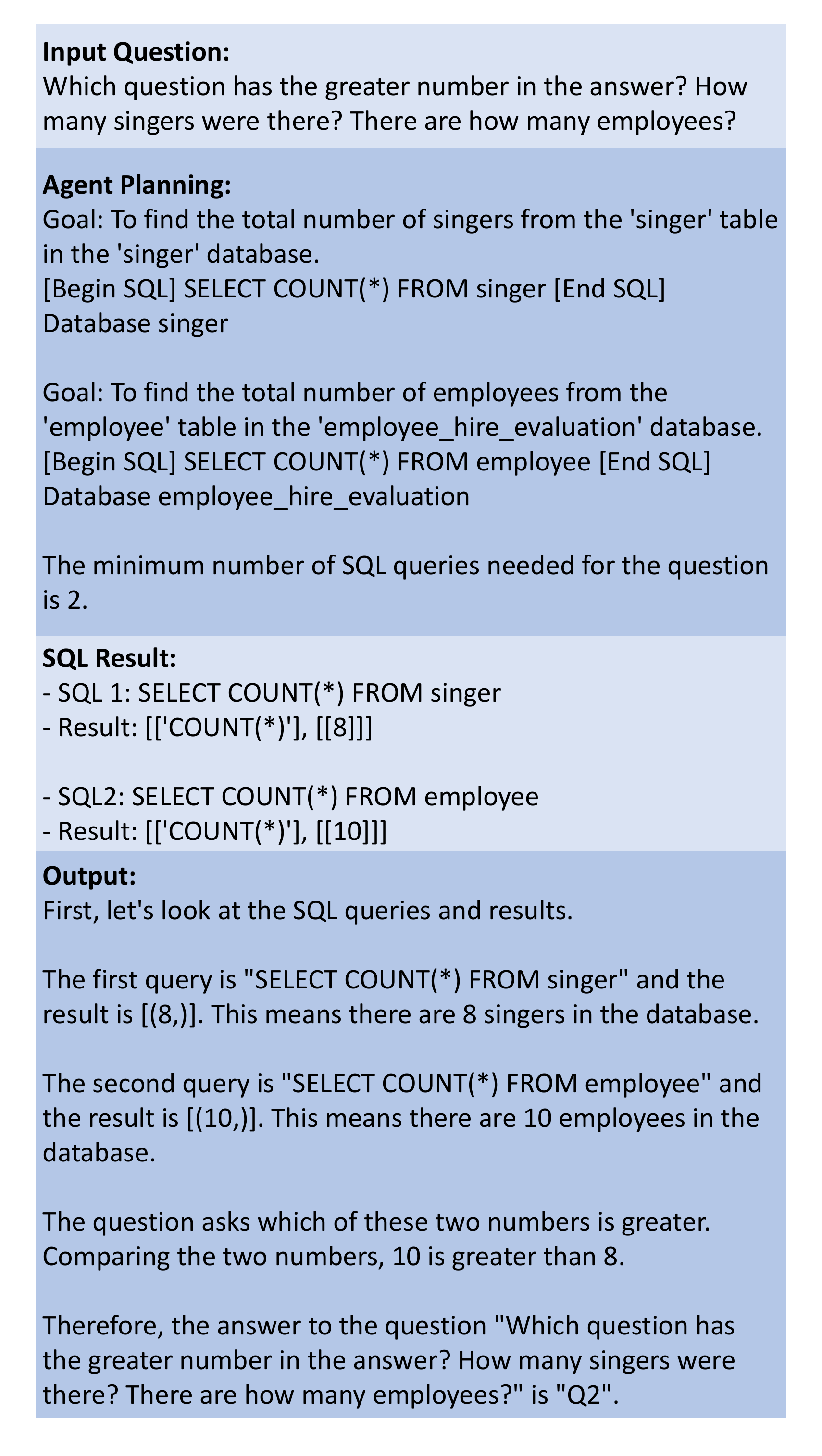}
    \caption{Sample case of the Claude-2 agent-enhanced model on a double database question.}
    \label{fig:case_study}
    \vspace{-3mm}
\end{figure}
}

\section{Conclusion}

In this work, we propose a novel framework to augment LLMs with relational databases. Experiments show that our method allows the LM to answer questions about database contents with reasonable accuracy. We believe our work would have a significant impact on various NLP applications in the real world.

\section{Limitations}

There are some limitations in this work. First, due to the lack of computing resources, we do not use open-source LLMs to evaluate our framework. Besides, the double DB questions in the experiments are artificial. It would be better if we could include real-world questions that require accessing multiple databases in our dataset. Third, this work does not consider the data privacy problem when letting LLMs access the database. It is a practical concern given the fact that the LLMs are usually run on the clouds.

\bibliography{reference}
\clearpage
\appendix

\section{Detailed Prompts}

\subsection{Context Switch Prompt}

Does context of the question contain answer for the question?Explain your decision.
If the decision is yes, output "(YES)". If not, output "(NO)"
For example:\\
Question: LA has 2 universities, SF has 3 universities. The number of universities in SF is {{?}}.\\
(YES) The number of universities in SF is explicitly given at position 30.\\
Question: The color of grass is {{?}}.\\
(NO) The context is empty\\
Question: LA has 2 universities. The number of universities in SF is {{?}}.\\
(NO) I cannot fine the number of universities in SF from the context.\\
Question: The number of Burger King in Vallejo city is 0. The number of Five guys in Vallejo city is {{?}}.\\
(NO) The context only provides the number of Burger King in Vallejo city, but not the number of Five guys in Vallejo city.\\
Question: The number of Burger King in Vallejo city is 0. The number of Five guys in Vallejo city is 1. Comparing the number of Burger King and Five guys in Vallejo city, we know {{?}} has more.\\
(YES) Although the context does not directly answer the question about which one has more, we can get the answer by comparing the number of Burger King and Five guys in Vallejo city, which is provided in the context.\\
Question: LA has 2 universities, SF has 3 universities. Comparing the number of universities in LA and SF, we know {{?}} has more universities.\\
(YES) The context provides information about the number of universities in both LA and SF, and the question directly asks about comparing the number of universities between the two cities to know which city has more universities.\\
Question: Beyonce Giselle Knowles-Carter (born September 4, 1981) is an American singer, songwriter, record producer and actress. Born and raised in Houston, Texas, she performed in various singing and dancing competitions as a child, and rose to fame in the late 1990s as lead singer of R\&B girl-group Destiny's Child. Managed by her father, Mathew Knowles, the group became one of the world's best-selling girl groups of all time. Their hiatus saw the release of Beyonce's debut album, Dangerously in Love (2003), which established her as a solo artist worldwide, earned five Grammy Awards and featured the Billboard Hot 100 number-one singles \"Crazy in Love\" and \"Baby Boy\". What was the name of Beyonce's first solo album?\\
(YES) The answer can be found at position 505 from the context string.\\
Question: The Legend of Zelda: Twilight Princess is an action-adventure game developed and published by Nintendo for the GameCube and Wii home video game consoles. It is the thirteenth installment in the The Legend of Zelda series. Originally planned for release on the GameCube in November 2005, Twilight Princess was delayed by Nintendo to allow its developers to refine the game, add more content, and port it to the Wii. The Wii version was released alongside the console in North America in November 2006, and in Japan, Europe, and Australia the following month. The GameCube version was released worldwide in December 2006.[b]" What year was the Legend of Zelda: Australian Princess originally planned for release?\\
(NO) The context does not mention Legend of Zelda: Australian Princess\\
Question: \{question\}\\

\subsection{Database Selection and Planning Prompt}

Based on the following schemas of selected databases and the input question.
Database {db}\\
Schema: ...\\
...\\
Input Question: \{question\}\\
What is the minimum number of SQL queries needed for the question? Generate the SQLs.\\
\#\#\# Use the following instructions to generate SQLs.\\
0) You can only use the databases provided above.\\
1) Generate each SQL one by one.\\
2) For each SQL, first explain its objective. The objective 
should contain all the details from the original question\\
3) Then generate the SQL query.\\
4) Finally output the database name.\\
5) Use the following format\\
Goal: xxx\\
*Begin SQL* select xxx from xxx *End SQL*\\
Database xxx\\
6) Generate all the SQL needed

\subsection{Value Correction Prompt}
The SQL below might contain errors. Try to correct it with the candidate values selected from the database and clarify the reasons. If the SQL does not need correction, directly return the original SQL.\\
\#\#\#\# Use the following instructions for fixing the SQL QUERY:\\
0) You can only change the string values in the SQL conditions.\\
1) Do not change the SQL structure\\
2) Use the database values that are explicitly mentioned in the Candidate Values.\\
3) Pay attention to the columns that are used for the JOIN by using the Foreign\_keys\\.
4) Use DESC and DISTINCT when needed.\\
5) Pay attention to the columns that are used for the GROUP BY statement.\\
6) Pay attention to the columns that are used for the SELECT statement.\\
7) Only change the GROUP BY clause when necessary (Avoid redundant columns in GROUP BY).\\
8) Use GROUP BY on one column only.\\
9) Do not use ANY or ALL.\\
\#\#\# Follow the format in the example.\\
Example:\\
    SQL: \\
    SELECT avg(RATING)  FROM RESTAURANT  JOIN GEOGRAPHIC ON RESTAURANT.CITY\_NAME = GEOGRAPHIC.CITY\_NAME  WHERE NAME = 'Tifft Jane Caterer'  AND RESTAURANT.CITY\_NAME = 'San Francisco City'\\
    Candidate Values:\\
    Table: restaurant; Column: name; Values: ['tifft jane caterer', "natty bumppo's", "wendy's old fashn hamburgers", "jennifer's bakery cafe", "mondtray's cafe", "marie callender's pie shop", "flintroy's bar-b-q", "monterey's fish house", "ruthie's taqueria", "wendy's old fashion hamburgers", "marie callender's pie shops", 'mama lupe taqueria', "ernie's neptune fish grotto", 'chubby jr burgers', "rebecca's mighty muffins", "flint's barbeque", 'pee wee muldoons', "bette's oceanview diner", "cybelle's gilman", "karlita's taco place"]\\
    Table: restaurant Column: city\_name; Values: ['san francisco', 'san fransisco', 'san jose', 'south san francisco', 'san carlos', 'san pablo', 'san juan bautista', 'san anselmo', 'san mateo', 'san lorenzo', 'san bruno', 'san martin', 'san ramon', 'santa cruz', 'san leandro', 'santa clara', 'saratoga', 'santa rosa', 'sausalito', 'pacific grove']\\
    Corrected SQL: \\
    SELECT AVG(RATING) FROM RESTAURANT JOIN GEOGRAPHIC ON RESTAURANT.CITY\_NAME = GEOGRAPHIC.CITY\_NAME WHERE NAME = 'tifft jane caterer' AND RESTAURANT.CITY\_NAME = 'san francisco';\\
   Reasons: In this corrected query, the values 'tifft jane caterer' and 'san francisco' are taken from the provided candidate values, preserving their case sensitivity.\\
SQL: ...\\
Candidate Values: (\{table\},\{column\},\{value\}), ...\\
Question: \{question\}
Corrected SQL:\\

\subsection{Error Message-based Self-Correction Prompt}

Database Schema:\\
...\\
Question: \{question\}\\
SQL: ...\\
Error Message: ...\\
The SQL is parsed based on the question and database schema. Correct it based on the error message.

\subsection{Return Type Decision Prompt}

Is the goal and SQL equivalent? Answer yes or no.\\
Question: \{question\}\\
SQL: \{SQL\}

\subsection{Output Generation Prompt}

\# SQL Results\\
...\\
Question: \{question\}\\
Based on the context, is there any SQL query and result helpful to answer the question?\\
If yes, utilize the information in SQL results to answer the question\\
If not, ignore the SQL and directly answer the question\\
Let's think step by step.

\section{Test Dataset Collection Process}
The questions are collected from four public datasets: (1) SQUAD-v2~\cite{rajpurkar2018know}; (2) Commonsense QA~\cite{talmor2018commonsenseqa}; (3) Spider~\cite{yu2018spider}; and (4) Dr-Spider~\cite{chang2023dr}. The first two datasets provide questions that do not require accessing databases, while the other two provide questions that require accessing one and only one database. In addition, we artificially generate questions that require accessing multiple databases. Specifically, we sample two questions from the Spider and Dr-Spider datasets. Both questions can be parsed into SQL queries that return a number. Then we combine the two questions with the template "Which question has a larger number as its answer. Q1: xxx; Q2: xxx". Although the question template is artificial, it requires the LLM to select the two correct databases, generate the correct retrieval target and SQL query for each database, and process the SQL results to answer the question. And since this type of questions only appears in the test stage, it can represent any real-world questions that require accessing multiple database questions.

\section{Additional Experiments}

\subsection{Context Switch}

Next, we show the effect of the context switch module on the overall agent design. Since the goal of context switch control is to avoid unnecessary retrieval, we evaluate its effect on the database selection to see if it could prevent false positive database retrieval.
Table \ref{tab:context_switch} shows the precision, recall, and f1-score of database selection where each question has 5 candidate databases. We can see that adding the context switch improves the precision by 0.05 without decreasing the recall. It suggests that the context switch successfully prevents redundant database retrieval and has no negative effect on the database selection recall. 

\begin{table}[]
\footnotesize
    \caption{Effect of context switch to the effectiveness of database selection}
    \label{tab:context_switch}
    \centering
    \begin{tabular}{c|c|c|c}\hlineB{3}
         & precision & recall & f1-score \\\hlineB{3}
    without Context Switch & 0.61 & \textbf{0.90} & 0.73\\
    with Context Switch & \textbf{0.66} & \textbf{0.90} & \textbf{0.76}\\\hlineB{3}
    \end{tabular}
\end{table}

\subsection{Case Study}

Figure \ref{fig:case_study} shows a sample case of the Claude-2 model on a double database question where 20 candidate databases are provided. The question asks the agent to compare the number of singers and the number of employees. Notably, the agent has to access two databases, "singer" and "employee\_hire\_evaluation" to obtain the actual values of the numbers. From the retrieval plan, we can see that the agent correctly understands the question and generates two retrieval goals to obtain the necessary data. And it correctly selects which databases are relevant and generates correct SQL queries. Furthermore, when provided with the SQL queries and results, the output generation module manages to extract information from them to generate the output. From this example, we can conclude that our agent has the ability to generate retrieval plans based on the provided databases and utilize the SQL results to generate outputs. However, we notice that in other double database questions, the LLM might make wrong answers when comparing the two numbers due to the inherent limitation of LLMs 

\begin{figure}
    \centering
    \includegraphics[width=0.95\linewidth,height=0.6\textheight]{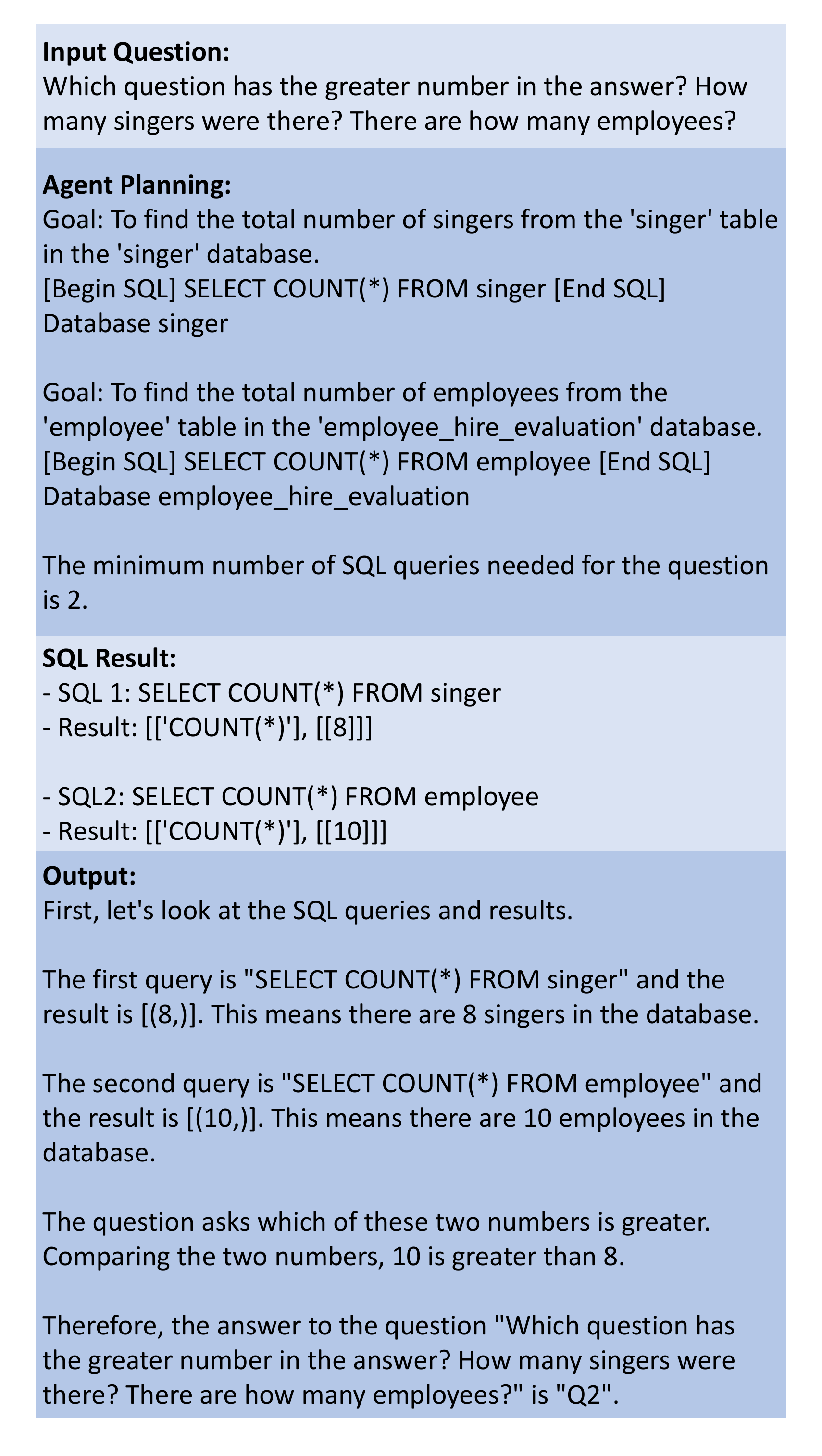}
    \caption{Sample case of the Claude-2 agent-enhanced model on a double database question.}
    \label{fig:case_study}
    \vspace{-3mm}
\end{figure}

\end{document}